\documentclass[aps,preprint,amsmath,amssymb,showpacs]{revtex4}

\usepackage{graphicx}
\begin{document}

\title{Effect of top quark spin on the unparticle couplings in $\gamma\gamma \to t\bar{t}$}

\author{\.{I}nan\c{c} \c{S}ahin}
\email[]{isahin@wisc.edu} \email[]{isahin@science.ankara.edu.tr}
\affiliation{Department of Physics, University of Wisconsin,
Madison, WI 53706, USA}
 \affiliation{Department of
Physics, Faculty of Sciences, Ankara University, 06100 Tandogan,
Ankara, Turkey}

\begin{abstract}
We investigate the potential of $\gamma\gamma$ collisions to probe
scalar unparticle couplings via top-antitop quark pair production.
We find 95$\%$ confidence level limits on the unparticle couplings
with an integrated luminosity of $500 fb^{-1}$ and $\sqrt{s}=1$ TeV
energy. We investigate the effect of top quark spin polarization on
the unparticle couplings. It is shown that spin polarization of the
top quark leads to a significant improvement in the sensitivity
limits.

\end{abstract}


\maketitle

\section{Introduction}
Scale invariance plays a crucial role in theoretical physics. A
possible scale invariant hidden sector that may interact weakly with
the Standard Model (SM) fields is being discussed intensively in the
literature. Based on a scale invariant theory by Banks-Zaks (BZ)
\cite{BZ}, Georgi proposed a new scenario \cite{Georgi, Georgi2} in
which SM fields and a scale invariant sector described by (BZ)
fields interact via the exchange of particles with a large mass
scale. At low energies this scale invariant sector manifests itself
as a non-integral number $d_U$ of particles called unparticles.
Several effective interaction terms between unparticles and SM
particles have been proposed and phenomenological \cite{Luo},
astrophysical and cosmological \cite{Davoudiasl} implications of
unparticles have been intensively studied in the literature.

The top quark possesses a large mass; its mass is at the electroweak
symmetry-breaking scale. Because of its large mass, top quark
 couplings are expected to be more sensitive to new physics than
other particles \cite{zhang}. Probing top quark couplings in the
context of new physics will be a crucial test of the SM. Top quark
couplings to unparticles have been analyzed in several papers
without taking into account of top quark spin polarization
\cite{Alan,Feng}. Because of its large mass the weak decay time for
the top quark is much shorter than the typical time for the strong
interactions to affect its spin \cite{bigi}. Therefore, the
information on its polarization is not disturbed by hadronization
effects but transferred to the decay products.  The effect of top
quark spin polarization on a possible new physics contribution from
extra dimensions and several effective Lagrangians have been widely
studied in the literature. Top spin analysis in the top-antitop pair
production processes has been performed regarding ADD and RS models
\cite{Smolek} and the effective Lagrangian approach \cite{Yuan}.

In this paper, we investigate the scalar unparticle contribution in
the process $\gamma\gamma \to t\bar{t}$. We take into account top
spin polarization along the direction of various spin bases to
improve the sensitivity bounds. We have shown that top spin
polarization leads to a significant improvement in the sensitivity
bounds.

In our calculations we consider the following effective interaction
operators between SM fields and scalar unparticles \cite{Cheung}:

\begin{eqnarray}
\frac{\lambda_S}{\Lambda_{U}^{d_{U}-1}}\bar{f}f O_U,\,\,\,\,\,
\frac{\lambda_{PS}}{\Lambda_{U}^{d_{U}-1}}\bar{f}i\gamma^5 f
O_U,\,\,\,\,\,\frac{\lambda_V}{\Lambda_{U}^{d_{U}}}\bar{f}\gamma^{\mu}f(\partial_{\mu}O_U)
,\,\,\,\,\,\frac{\kappa}{\Lambda_{U}^{d_{U}}}
G_{\mu\nu}G^{\mu\nu}O_U
\end{eqnarray}
Here $f$ stands for a SM fermion and $G^{\mu\nu}$ denotes the gauge
field strength. Feynman rules for these operators have been given in
\cite{Cheung}. The vertex functions generated from operators (1) are
given by

\begin{eqnarray}
i\frac{\lambda_S}{\Lambda_{U}^{d_{U}-1}},\,\,\,\,\,
-\frac{\lambda_{PS}}{\Lambda_{U}^{d_{U}-1}}\gamma^5,\,\,\,\,\,
\frac{\lambda_V}{\Lambda_{U}^{d_{U}}}\gamma^{\mu}p_{\mu},\,\,\,\,\,
4i\frac{\kappa}{\Lambda_{U}^{d_{U}}}(-p_1\cdot p_2
g^{\mu\nu}+p_1^{\nu}p_2^{\mu}),\,\,\,\,\,
\end{eqnarray}
where $p$ is the unparticle momentum and $p_1$ and $p_2$ are the
momenta of two photons. For convention, we assume that all the
momenta are incoming to the vertex. The vertex
$\frac{\lambda_V}{\Lambda_{U}^{d_{U}}}\gamma^{\mu}p_{\mu}$ does not
contribute to the process when the unparticle couples to a
on-mass-shell fermion current. In some papers, it is assumed for
simplicity that the coupling constants are all equal. To carry out a
more general treatment we assume that they are different and
distinguished by additional labels.

\section{Spin dependent cross section for $t\bar t$ production}

The research and development on linear $e^{+}e^{-}$ colliders have
been progressing and the physics potential of these future machines
is under study. After linear colliders are constructed its operating
modes of $e\gamma$ and $\gamma\gamma$ are expected to be designed
\cite{akerlof}. A real gamma beam is obtained through Compton
backscattering of laser light off linear electron beam where most of
the photons are produced at the high energy region. The luminosities
for $e\gamma$ and $\gamma\gamma$ collisions turn out to be of the
same order as the one for $e^{+}e^{-}$ \cite{Ginzburg}, so the cross
sections for photoproduction processes with real photons are
considerably larger than virtual photon case.

The spectrum of the backscattered photons is given by
\cite{Ginzburg}

\begin{eqnarray}
f_{\gamma/e}(y)={{1}\over{g(\zeta)}}[1-y+{{1}\over{1-y}}
-{{4y}\over{\zeta(1-y)}}+{{4y^{2}}\over {\zeta^{2}(1-y)^{2}}}],
\end{eqnarray}

where

\begin{eqnarray}
g(\zeta)=&&(1-{{4}\over{\zeta}}
-{{8}\over{\zeta^{2}}})\ln{(\zeta+1)}
+{{1}\over{2}}+{{8}\over{\zeta}}-{{1}\over{2(\zeta+1)^{2}}},
\end{eqnarray}
with $\zeta=4E_{e}E_{0}/M_{e}^{2}$. $E_{0}$ is the energy of the
initial laser photon and $E_{e}$ is the energy of the initial
electron beam before Compton backscattering. $y$ is the fraction
which represents the ratio between the scattered photon and the
initial electron energy for the backscattered photons moving along
the initial electron direction. The maximum value of $y$ reaches
0.83 when $\zeta=4.8$ in which the backscattered photon energy is
maximized without spoiling the luminosity.

The integrated cross section for $t\bar t$ production via
$\gamma\gamma$ fusion can be obtained by the following integration:

\begin{eqnarray}
d\sigma(e^+e^- \to \gamma\gamma \to t\bar
t)=\int_{z_{min}}^{z_{max}}dz \,2z \, d\hat{\sigma}(\gamma\gamma \to
t\bar t) \, \int_{z^2/y_{max}}^{y_{max}}
\frac{dy}{y}f_{\gamma/e}(y)f_{\gamma/e}(z^2/y)
\end{eqnarray}
where $d\hat{\sigma}(\gamma\gamma \to t\bar t)$ is the cross section
of the subprocess and the center of mass energy of the $e^+e^-$
system $\sqrt s$, is related to the center of mass energy of the
$\gamma\gamma$ system $\sqrt{\hat s}$, by $\hat s =z^2s$.

In the presence of the couplings (2), $\gamma\gamma \to t\bar t$
scattering is described by three tree-level diagrams. One can see
from Fig.1 that the s-channel diagram contains unparticle exchange
and modifies the SM amplitudes.

Since the top quark is very heavy, its helicity is frame dependent
and changes under a boost from one frame to another. The helicity
and chirality states do not coincide with each other and there is no
reason to believe that the helicity basis will give the best
description of the spin of the top quarks. Therefore, it is
reasonable to study other spin bases better than helicity for the
top quark spin. The spin four-vector of a top quark is defined by

\begin{eqnarray}
s_{t}^{\mu}=(\frac{\vec{p}_{t}\cdot \vec{s^{\prime}}}{m_{t}} \,,\,
\vec{s^{\prime}}+\frac{\vec{p}_{t}\cdot \vec{s^{\prime}}
}{m_{t}(E_{t}+m_{t})}\vec{p}_{t})
\end{eqnarray}
where $(s_{t}^{\mu})_{RF}=(0,\vec{{s}^{\prime}})$ in the top quark
rest frame. The laboratory frame is the $e^{+}e^{-}$ center of mass
system where the cross section is performed. $\vec{{s}^{\prime}}$
should be obtained by a Lorentz boost from the laboratory frame:

\begin{eqnarray}
  \vec{{s}^{\prime}}=\lambda \frac{\vec{{p}^{\star}}}
  {|\vec{{p}^{\star}}|} ,\,\,\,\,\, \lambda=\pm 1.\nonumber\\
\vec{{p}^{\star}}=\vec{p}+\frac{\gamma-1}{\beta^{2}}
(\vec{\beta}\cdot \vec{p})\vec{\beta}
 -E\gamma \vec{\beta}
\end{eqnarray}
Here $\vec{p}$  is the momentum of the particle moving along the top
spin direction in the laboratory frame and $\vec{{p}^{\star}}$ is
the momentum observed in the rest frame of the top quark. A similar
treatment can also be done for antitop quarks.

We consider two different top spin directions in the laboratory
frame; one of the incoming photon beam direction and the helicity
basis. We have calculated the polarized cross sections for the above
spin directions of the top and unpolarized antitop quark.
Calculations for polarized antitop and unpolarized top quark would
lead to similar results. Phase space integrations have been taken by
a Monte Carlo routine. In the cross section calculations we have
performed a boost to obtain $\vec{{p}^{\star}}$ at each point in
phase space.

Spin dependent SM squared amplitudes are given by

\begin{eqnarray}
|M_1|^2=\frac{4\,g_e^4}{27(q_1^2-m_t^2)^2}&&\left\{8\,k_1\cdot p_t\,
k_1\cdot p_{\bar t}+16\,m_t^2\,k_1\cdot p_t+8\,m_t^2\,k_1\cdot
p_{\bar t}-16\,m_t^4-8\,m_t^2\,k_1\cdot p_t\,s_t \cdot s_{\bar
t}\right.\nonumber\\&&\left.+8\,m_t^4\,s_t \cdot s_{\bar
t}-8\,m_t^2\,s_t\cdot k_1\, s_{\bar t}\cdot k_1+8\,m_t^2\,s_t\cdot
k_1 \, s_{\bar t}\cdot
p_t-8\,m_t^2\,p_t\cdot p_{\bar t}\right\}\\
\nonumber \\ |M_2|^2=&&|M_1|^2\,\,\,(k_1 \longleftrightarrow k_2)\\
\nonumber
\end{eqnarray}
\begin{eqnarray}
2\,Re(M_1^{\dag}M_2)&&=\frac{32\,g_e^4}{27(q_1^2-m_t^2)(q_2^2-m_t^2)}\left\{-m_t^2\,k_1\cdot
p_{\bar t} +2 m_t^2\,k_2\cdot p_t -m_t^2\,k_2\cdot p_{\bar
t}-m_t^4\right.\nonumber \\
&&\left.-k_1\cdot p_{\bar t}\,k_2\cdot p_t \, s_t\cdot s_{\bar
t}-m_t^2\,k_2\cdot p_t \,s_t\cdot s_{\bar t}+m_t^4\,s_t\cdot s_{\bar
t}+k_2\cdot p_t\,s_t\cdot p_{\bar t}\,s_{\bar t}\cdot k_1 \right.\nonumber \\
&&\left.+k_2\cdot p_{\bar t}\,s_t\cdot k_1\,s_{\bar t}\cdot p_t
-m_t^2\,s_t\cdot k_1\,s_{\bar t}\cdot p_t+k_1\cdot p_{\bar
t}\,s_t\cdot k_2\,s_{\bar t}\cdot p_t-m_t^2\,s_t\cdot k_2\,s_{\bar
t}\cdot p_t\right.\nonumber \\
&&\left.+m_t^2\,s_t\cdot p_{\bar t}\,s_{\bar t}\cdot p_t+2\,k_2\cdot
p_t\,p_t\cdot p_{\bar t}+m_t^2\,p_t\cdot p_{\bar t}-m_t^2\,s_t\cdot
s_{\bar t}\,p_t\cdot p_{\bar t}-s_t\cdot k_2\,s_{\bar t}\cdot
k_1\,\right.\nonumber
\\&&\left. \times p_t\cdot p_{\bar t}-s_t\cdot k_1\,s_{\bar t}\cdot k_2\,p_t\cdot p_{\bar t}
+k_1\cdot p_t(2m_t^2-k_2\cdot p_{\bar t}\,s_t\cdot s_{\bar
t}-m_t^2\,s_t\cdot s_{\bar t}+s_t\cdot p_{\bar t}\,s_{\bar t}\cdot
k_2\right.\nonumber \\
&&\left.+2\,p_t\cdot p_{\bar t})+k_1\cdot k_2(m_t^2(-1+2\,s_t\cdot
s_{\bar t})-s_t\cdot p_{\bar t}\,s_{\bar t}\cdot p_t+(-2+s_t\cdot
s_{\bar t})p_t\cdot p_{\bar t})\right\}
\end{eqnarray}
where $k_1$ and $k_2$ are the momenta of incoming photons, $p_t$ and
$p_{\bar t}$ are the momenta of outgoing top and antitop quarks and
$s_t^\mu$ and $s_{\bar t}^\mu$ are the spin four-vectors of top and
antitop quarks. The u-channel amplitude can be obtained from the
t-channel by interchanging the incoming photon momenta
($k_1\longleftrightarrow k_2$).

The spin dependent unparticle contributions are given by the
following amplitudes:

\begin{eqnarray}
|M_3|^2=&&\frac{3\,A_{d_U}^2}{\sin^2(\pi
d_U)}\,|q_3^2|^{2d_U-4}\left(\frac{\kappa^2}{\Lambda_U^{4d_U-2}}\right)\left\{\lambda_{S}^2\,\left[2\,(k_1\cdot
 k_2)^2\,(-m_t^2+s_t\cdot p_{\bar t}\,\,s_{\bar t}\cdot p_t\right.\right.\nonumber \\
 &&\left.\left.+p_t\cdot p_{\bar t}\,+
 s_t\cdot s_{\bar t}\,(m_t^2-p_t\cdot p_{\bar t}))\right]+\lambda_{PS}^2 \left[2\,(k_1\cdot
 k_2)^2\,(m_t^2-s_t\cdot p_{\bar t}\,\,s_{\bar t}\cdot p_t\right.\right.\nonumber \\
 &&\left.\left.+p_t\cdot p_{\bar t}
 +s_t\cdot s_{\bar t}\,(m_t^2+p_t\cdot p_{\bar t}))\right]\right\},
\end{eqnarray}

\begin{eqnarray}
2\,Re(M_1^{\dag}M_3)=&&-\frac{8\,g_e^2\,A_{d_U}|q_3^2|^{d_U-2}}{3\,\sin(\pi
d_U)}\,\,\frac{1}{(q_1^2-m_t^2)}\,\left(\frac{\kappa}{\Lambda_U^{2d_U-1}}\right)
\nonumber \\&&\times \left\{\lambda_S\left[m_t(k_2\cdot p_t(k_1\cdot
p_t-k_1\cdot p_{\bar t}-(k_1\cdot p_t-k_1\cdot p_{\bar t})s_t\cdot
s_{\bar
t}\right.\right.\nonumber \\
&&\left.\left.-s_t\cdot p_{\bar t}\,s_{\bar t}\cdot k_1+s_t\cdot
k_1\,s_{\bar t}\cdot p_t)+k_1\cdot k_2(s_t\cdot p_{\bar t}\,s_{\bar
t}\cdot k_1-s_t\cdot k_1\, s_{\bar t}\cdot
p_t\right.\right.\nonumber \\
&&\left.\left.+s_t\cdot p_{\bar t}\,s_{\bar t}\cdot p_t+s_t\cdot
s_{\bar t}\,(k_1\cdot p_t-k_1\cdot p_{\bar t}+m_t^2-p_t\cdot p_{\bar
t})-k_1\cdot p_t+k_1\cdot p_{\bar
t}\right.\right.\nonumber \\
&&\left.\left.-m_t^2+p_t\cdot p_{\bar t}))\cos(\pi d_U)-((k_1\cdot
k_2-k_2\cdot p_t)(-1+s_t\cdot s_{\bar t})\epsilon^{k_1\,p_t\,p_{\bar
t}\,s_{\bar
t}}\right.\right.\nonumber \\
&&\left.\left.+k_1\cdot k_2(s_{\bar t}\cdot
p_t\,\epsilon^{k_1\,p_{\bar t}\,s_t\,s_{\bar t}}-s_{\bar t}\cdot
k_1\,\epsilon^{p_t\,p_{\bar t}\,s_t\,s_{\bar t}})+k_2\cdot
p_t(-s_{\bar t}\cdot p_t\,\epsilon^{k_1\,p_{\bar t}\,s_t\,s_{\bar
t}}\right.\right.\nonumber \\
&&\left.\left.+s_{\bar t}\cdot k_1\,\epsilon^{p_t\,p_{\bar
t}\,s_t\,s_{\bar t}}))\sin(\pi
d_U)\right]+\lambda_{PS}\left[m_t\cos(\pi d_U)(k_2\cdot
p_t(\epsilon^{k_1\,p_t\,s_t\,s_{\bar
t}}\right.\right.\nonumber \\
&&\left.\left.+\epsilon^{k_1\,p_{\bar t}\,s_t\,s_{\bar t}})-k_1\cdot
k_2(\epsilon^{k_1\,p_t\,s_t\,s_{\bar t}}+\epsilon^{k_1\,p_{\bar
t}\,s_t\,s_{\bar t}}+\epsilon^{p_t\,p_{\bar t}\,s_t\,s_{\bar
t}}))-(k_1\cdot
k_2\right.\right.\nonumber \\
&&\left.\left.\times(-m_t^2\,s_t\cdot k_1+k_1\cdot p_t\,s_t\cdot
p_{\bar t}+m_t^2\,s_t\cdot p_{\bar t}+m_t^2\,s_{\bar t}\cdot
k_1-k_1\cdot p_{\bar t}\,s_{\bar t}\cdot p_t\right.\right.\nonumber \\
&&\left.\left.+m_t^2\,s_{\bar t}\cdot p_t-s_t\cdot k_1\,p_t\cdot
p_{\bar t}+s_{\bar t}\cdot k_1\,p_t\cdot p_{\bar t})+k_2\cdot
p_t(-k_1\cdot p_t\,s_t\cdot p_{\bar t}
\right.\right.\nonumber \\
&&\left.\left.-m_t^2\,s_{\bar t}\cdot k_1+k_1\cdot p_{\bar
t}\,s_{\bar t}\cdot p_t-s_{\bar t}\cdot k_1\,p_t\cdot p_{\bar
t}+s_t\cdot k_1(m_t^2+p_t\cdot p_{\bar t})))\right.\right.\nonumber \\
&&\left.\left.\times \sin(\pi d_U)\right]\right\},\\
\nonumber \\
 2\,Re(M_2^{\dag}M_3)=&&2\,Re(M_1^{\dag}M_3)\,\,(k_1
\longleftrightarrow k_2),
\end{eqnarray}
where $A_{d_{U}}\equiv\frac{16\pi^{\frac{5}{2}}}{(2\pi)^{2d_{U}}}
\frac{\Gamma(d_{U}+\frac{1}{2})}{\Gamma(d_{U}-1)\Gamma(2d_{U})}$.
The trigonometric functions  $\cos(\pi d_U)$ and $\sin(\pi d_U)$ in
the amplitudes (12) and (13) originate from the complex phase
associated with the s-channel propagator and may lead to interesting
interference effects with the standard model amplitudes.

The integrated cross section as a function of the unparticle
couplings $\kappa$, $\lambda_S$ and $\lambda_{PS}$ for various top
quark spin polarizations are plotted in Figs. \ref{fig2} -
\ref{fig6}. In the figures the center of mass energy of the $e^+
e^-$ system is $\sqrt{s}$=1 TeV. $\sigma$ versus $\kappa$ graphs are
plotted under the assumption that $\lambda_S$=$\lambda_{PS}$=1. In
the graphs of $\sigma$ versus $\lambda_S$ and $\sigma$ versus
$\lambda_{PS}$, $\kappa$ is taken to be 1; $\lambda_{PS}$ and
$\lambda_{S}$ are taken to be zero, respectively. We see from these
figures that the deviation of the cross section from its SM value
increases with decreasing $d_U$. This is very clear from the factors
$\frac{1}{\Lambda_U^{4d_U-2}}$ and $\frac{1}{\Lambda_U^{2d_U-1}}$ in
the squared amplitudes (11)-(13). The influence of spin polarization
on the behavior of the cross sections can be observed from the
figures.

\section{Angular Correlations Between Top Decay Products and Sensitivity limits on the unparticle couplings}

Top quark spin polarization can be determined from the angular
distributions of its decay products. We consider the dominant decay
chain of the top quark to leptons in the standard model, $t \to
W^{+}b(W^{+} \to l^{+}\nu)$. The differential cross section for the
complete process including subsequent top decay is given by

\begin{eqnarray}
d\sigma \left(\gamma \gamma \to t\bar{t} \to b \,\ell^{+}
\nu_{\ell}\,\bar{t}\,
\right)=&&\frac{1}{2s}|M|^{2}\frac{d^{3}p_{1}}{(2\pi)^{3}2E_{1}}\frac{d^{3}p_{2}}{(2\pi)^{3}2E_{2}}
\frac{d^{3}p_{3}}{(2\pi)^{3}2E_{3}}\frac{d^{3}p_{\bar
t}}{(2\pi)^{3}2E_{\bar t}}
\nonumber \\
&&\times(2\pi)^{4}\delta^{4}\left(k_1+k_2-p_1-p_2-p_3-p_{\bar t}
\right),
\end{eqnarray}
where $k_1$ and $k_2$ are the momenta of the incoming photons,
$p_1$, $p_2$ and $p_3$  are the momenta of the outgoing fermions and
$p_{\bar t}$ is the momentum of the outgoing antitop quark.
$|M|^{2}$ is the square of the full amplitude which is averaged over
the initial spins and summed over the final spins. The full
amplitude can be expressed as follows:

\begin{eqnarray}
|M|^{2}(2\pi)^{4}\delta^{4}\left(k_1+k_2-p_1-p_2-p_3-p_{\bar t}
\right)=&&\int\frac{d^{4}q}{(2\pi)^{4}}
\left|\sum_{s_{t}}M_{a}(s_{t})D_{t}(q^{2})M_{b}(s_{t})\right|^{2}\nonumber
\\
&&\times(2\pi)^{4}\delta^{4}\left(k_{1}+k_{2}-p_{\bar
t}-q\right)\nonumber
\\ &&\times(2\pi)^{4}\delta^{4}\left(q-p_{1}-p_{2}-p_{3}\right),
\end{eqnarray}
where q and $s_{t}$ are the internal momentum and spin of the top
quark. $D_{t}(q^{2})$ is the Breit-Wigner propagator factor.
$M_{a}(s_{t})$ is the amplitude for the process $\gamma\gamma \to
t\bar{t}$ with on shell t quark. $M_{b}(s_{t})$ is the decay
amplitude for $t \to b \ell^{+} \nu_{\ell}$. The square of the decay
amplitude summed over the final fermion spins is given by

\begin{eqnarray}
|M_{b}(s_{t})|^{2}=\frac{2g_{w}^{4}}{[(p_{t}-p_{b})^{2}-m_{w}^{2}]^{2}}(p_{b}\cdot
p_{t}-p_{b}\cdot p_{\ell})(p_{\ell}\cdot p_{t}-m_{t}(s_{t}\cdot
p_{\ell}))
\end{eqnarray}
Here $p_\ell$ and $p_b$ are the momenta of the final lepton and b
quark.

It is easy to show that interference terms from different spin
states will vanish after integrating the decay part in (15) over
azimuthal angles of the top quark decay products. Then the full
cross section can be written as a product of production and decay
parts. Using the narrow width approximation and after some simple
algebra, the following result can be obtained:

\begin{eqnarray}
d\sigma\left(\gamma \gamma \to t\bar{t} \to b \,\ell^{+}
\nu_{\ell}\,\bar{t}\,\right)=\left[d\sigma\left(\gamma \gamma \to
 \uparrow t\,\bar{t}\,\right)\frac{d\Gamma\left(\uparrow t \to b \ell^{+}
\nu_{\ell}\right)}{\Gamma\left(t \to b \ell^{+} \nu_{\ell}\right)}
\right. \nonumber \\ \left. +d\sigma\left(\gamma \gamma \to
 \downarrow t\,\bar{t}\,\right)\frac{d\Gamma\left(\downarrow t \to b \ell^{+}
\nu_{\ell}\right)}{\Gamma\left(t \to b \ell^{+}
\nu_{\ell}\right)}\right]BR\left(t \to b \ell^{+} \nu_{\ell}\right),
\end{eqnarray}
where $BR\left(t \to b \ell^{+} \nu_{\ell}\right)$ is the leptonic
branching ratio for the top quark. Up and down arrows indicate spin
up and spin down cases along a specified spin quantization axis
respectively. $d\Gamma \left(\uparrow t \to b \ell^{+}
\nu_{\ell}\right)$ and $d\Gamma \left(\downarrow t \to b \ell^{+}
\nu_{\ell}\right)$ are the differential decay rates for the
polarized top quarks. The unpolarized rate is given by; $d\Gamma
\left(t \to b \ell^{+} \nu_{\ell}\right)=d\Gamma \left(\uparrow t
\to b \ell^{+} \nu_{\ell}\right)+d\Gamma \left(\downarrow t \to b
\ell^{+} \nu_{\ell}\right)$. For a fixed top quark spin, polarized
production cross sections can be obtained from a fit to the polar
angle distribution of the top decay product, an outgoing charged
lepton in the top rest frame. To be precise, the polarized
production cross sections $d\sigma\left(\gamma \gamma \to \uparrow
t\,\bar{t}\,\right)$ or
 $d\sigma\left(\gamma \gamma \to\downarrow t\,\bar{t}\,\right)$
can be obtained from a fit to the $d\Gamma\left(\uparrow t \to b
\ell^{+} \nu_{\ell}\right)$ or $d\Gamma\left(\downarrow t \to b
\ell^{+} \nu_{\ell}\right)$ distributions (17).

It is important to reconstruct the rest frame of the top quark,
since this is essential for its polarization identification. Since
it is impossible to detect a neutrino from a decaying top quark, it
is difficult to reconstruct the top quark momentum. On the other
hand the top momentum can be reconstructed from the antitop momentum
and the momenta of the incoming photons.  We assume that the antitop
quark is unpolarized; therefore, semi-leptonic decay is not a
necessary assumption for the antitop quark. Hence it is reliable to
assume that the antitop momentum is reconstructable. Reconstruction
of incoming photon momenta can be achieved by means of very forward
detectors in the ILC. Detection of the Compton backscattered
electron and positron, scattered at an almost zero-degree angle, in
the very forward detectors allow us to reconstruct the Compton
backscattered photon momenta. This kind of measurement was performed
by the H1 collaboration at HERA, but for Weizsacker-Williams photons
from the electron or proton \cite{H1}. In the LHC, it is foreseen
that one may equip two LHC experiments, ATLAS and CMS, with very
forward detectors which allow one to detect intact scattered protons
at very small angles after the collision. Therefore, it is
reasonable to assume a similar very forward detector equipment for
the ILC \cite{LHCphoton}.

As a concrete result; we have obtained 95$\%$ confidence level
limits on the unparticle couplings $\kappa$, $\lambda_S$ and
$\lambda_{PS}$ using a simple $\chi^{2}$ analysis at $\sqrt s$=1 TeV
and an integrated luminosity $L_{int}=500 fb^{-1}$ without
systematic errors. The $\chi^{2}$ function is given by

\begin{eqnarray}
\chi^{2}=\left(\frac{\sigma_{SM}-\sigma(\kappa,\lambda_{S},\lambda_{PS})}{\sigma_{SM}
\,\, \delta}\right)^{2},
\end{eqnarray}
where $\delta=\frac{1}{\sqrt{N}}$ is the statistical error. The
expected number of events has been calculated considering the
leptonic decay channel of the W boson as the signal $N=AL_{int}B(t
\to W^{+}b(W^{+} \to l^{+}\nu))B(\bar t \to W^{-}\bar b (W^{-} \to
l^{-}\bar{\nu}))\,\sigma$, where $A$ is the overall acceptance.

The limits on the unparticle couplings are given in Tables
\ref{tab1}-\ref{tab3} with an acceptance of $A=0.7$. This acceptance
value is assumed for the b-tagging efficiency as in references
\cite{btagging}. On the other hand, using an acceptance of 0.5
instead of 0.7 does not spoil our limits more than by a factor of
1.17. In Table \ref{tab1} we consider unpolarized top quark and in
Tables \ref{tab2} and \ref{tab3} we consider various top quark spin
polarizations. We omit the limits on the scalar coupling $\lambda_S$
for the polarized top quark, since $\lambda_S$ is insensitive to the
polarization. Limits on $\kappa$ have been calculated under the
assumption that $\lambda_S$=$\lambda_{PS}$=1, and limits on
$\lambda_S$ and $\lambda_{PS}$ have been calculated under the
assumption that $\kappa$=1 and the other remaining coupling is zero.
We see from the tables that significant improvements are obtained in
the sensitivity bounds by taking into account top quark spin
polarization for large values of the scale dimension $d_U\in[1,2]$.
For instance, the helicity right spin polarization configuration of
the top improves the lower bound on $\kappa$ by the factors of 7.5,
12.8 and 12 for $d_U$=1.5, $d_U$=1.7 and $d_U$=1.9, respectively.
Helicity right improves the lower bound, and helicity left improves
the upper bound on $\lambda_{PS}$ by a same factor of 4 for
$d_U$=1.7 . These improvement factors are 2.3, 3.8 and 3.2 for
$d_U$=1.3, $d_U$=1.5 and $d_U$=1.9. It is observed from the tables
that the improvement factors decrease as $d_U$ decreases and
approaches to 1. This behavior is reasonable from the analytic
expressions of the interference terms (12) and (13). The
interference of unparticle contributions with the SM amplitudes
explicitly depends on the scale dimension $d_U$ via its
trigonometric functions. Therefore any variation in $d_U$  should
produce some interference effects like the one which we have
observed from the tables. We see from expressions (12) and (13) that
interference terms are odd functions of the unparticle couplings.
Therefore, these interference effects spoil the symmetric behavior
of the cross section in the negative and positive intervals of the
unparticle couplings. Of course, it is very difficult to predict the
effect of an interference contribution from the analytic expressions
(12) and (13) precisely. But numerical results show that the
improvement provided by the polarization generally increases as
$d_U$ increases for $\kappa$ and $\lambda_{PS}$. We observe from
Figs.\ref{fig2}-\ref{fig6} that deviation of the cross section from
the SM for $d_U$=1.5 depends significantly on the polarization for
$\kappa$ and $\lambda_{PS}$. Altering the spin orientations, spin up
to down or helicity right to left, significantly changes the
behavior of the cross sections as a function of the couplings
$\kappa$ or $\lambda_{PS}$ for $d_U$=1.5. This is very compatible
with the results in the tables.

Another useful quantity about spin-induced angular correlations, one
that may be sensitive to new physics is the spin asymmetry,

\begin{eqnarray}
 A_{\uparrow\downarrow}=\frac{N_{\uparrow}-N_{\downarrow}}{N_{\uparrow}+N_{\downarrow}}
\end{eqnarray}
Here the subscript up arrow $\uparrow$ (down arrow $\downarrow$)
stands for spin up (spin down), and $N$ represents the number of
events for the corresponding spin. Substituting the spin dependent
amplitude square (16), into (17) and after some algebra we can reach
the following expression:

\begin{eqnarray}
\label{asymm}
 \frac{1}{\sigma_{T}}\frac{d\sigma}{dcos\theta}=\frac{1}{2}(1+A_{\uparrow\downarrow}cos\theta),
\end{eqnarray}
where $\theta$ is defined as the angle between the charged lepton
(from the decaying top quark) and the top quark spin quantization
axis in the rest frame of the top quark.

Spin asymmetry is a significant observable about spin-induced
angular correlations. The polarization of the top quark can be
extracted from the $cos \theta$ distribution in (\ref{asymm}).  We
see from (\ref{asymm}) that the up or down polarization state of the
top quark is determined by the sign of  $cos\theta$. The size of the
spin asymmetry shows how we can easily observe angular correlations
and therefore the polarized cross sections.

In table \ref{tab4} we have presented spin asymmetries for the
helicity basis and one of the incoming photon beam direction. In the
table, the unparticle couplings are taken to be $\kappa$=1,
$\lambda_S$=0 and $\lambda_{PS}$=1. Spin asymmetries for $\kappa$=1,
$\lambda_S$=1 and $\lambda_{PS}$=0 are all zero, so we have not
presented them in the table. The zero spin asymmetry for $\kappa$=1,
$\lambda_S$=1 and $\lambda_{PS}$=0 exhibits that the scalar coupling
$\lambda_S$ is insensitive to the polarization. On the other hand,
this information provides us the opportunity to dissociate
$\lambda_{PS}$ from $\lambda_S$ by studying the top spin
asymmetries.

\section{Conclusions}

We have investigated the potential of $\gamma\gamma \to t\bar{t}$
with polarized top quarks to probe the scalar unparticle top and
scalar unparticle photon couplings. The most sensitive results are
obtained for $d_U$=1.1. The sensitivity limits get worse as the
$d_U$ increases. This is clear from the inverse powers of the energy
scale $\Lambda_U$ and common in most processes. On the other hand,
top spin polarization leads to a significant improvement to the
sensitivity limits for large values of  $d_U\in[1,2]$.

We have also investigated the effect of the unparticle couplings on
the spin asymmetries. We show that the spin asymmetries are
insensitive to the scalar coupling  $\lambda_S$. On the other hand,
they are sensitive to the pseudoscalar coupling $\lambda_{PS}$, and
the spin asymmetries are large for large values of  $d_U$.
Therefore, at least in principle, it is possible to dissociate
$\lambda_{PS}$ from $\lambda_S$  by measuring the top spin asymmetry
in $\gamma\gamma \to t\bar{t}$.

\begin{acknowledgments}
The author acknowledges support through the Scientific and Technical
Research Council (TUBITAK) BIDEB-2219 grant.
\end{acknowledgments}


\pagebreak

\begin{figure}
\includegraphics{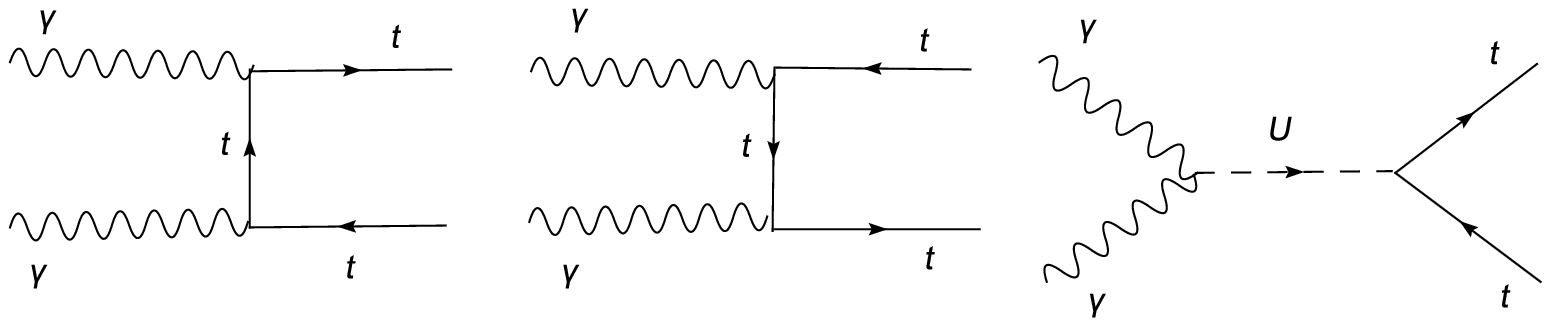}
\caption{Tree-level Feynman diagrams for $\gamma\gamma \to t\bar t$
in the presence of scalar unparticle couplings.\label{fig1}}
\end{figure}

\begin{figure}
\includegraphics{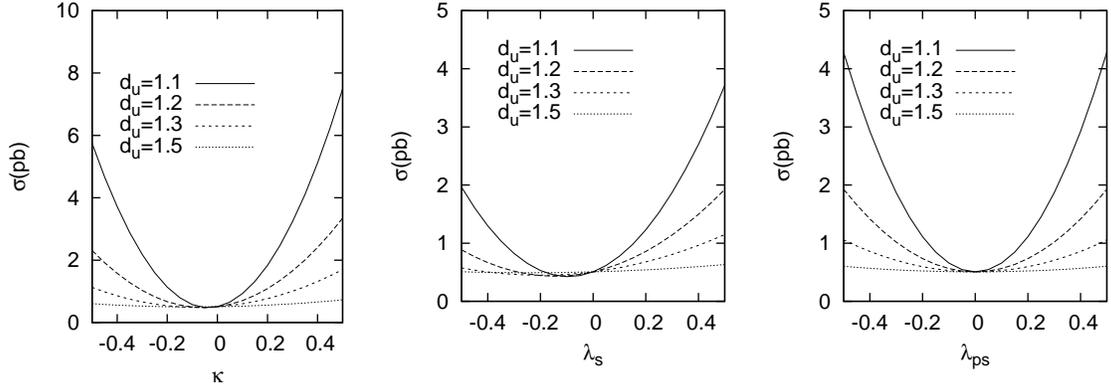}
\caption{ The integrated cross section of $\gamma\gamma \to
t\bar{t}$ as a function of the unparticle couplings $\kappa$,
$\lambda_S$ and $\lambda_{PS}$ for an unpolarized top-antitop quark
pair. Legends are for various values of the scale dimension $d_{U}$
and $\sqrt{s}=\Lambda_{U}$=1 TeV. \label{fig2}}
\end{figure}

\begin{figure}
\includegraphics{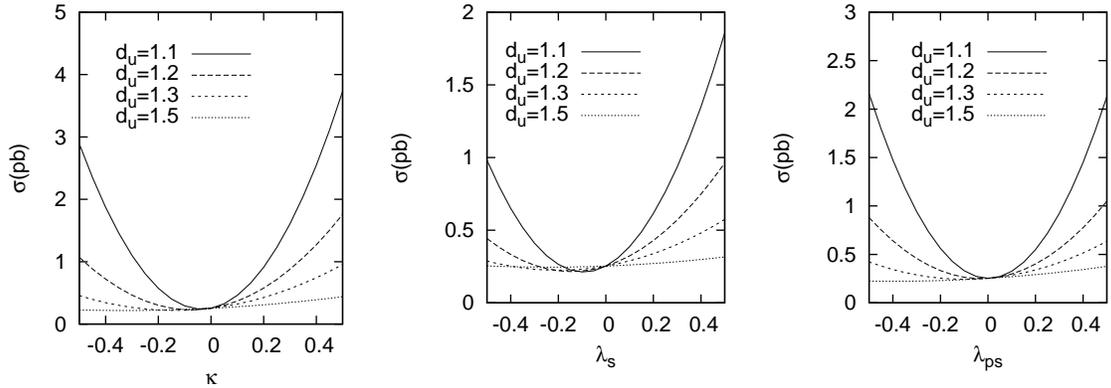}
\caption{ The same as Fig.2, but the top quark is in the helicity
basis with left helicity and the antitop quark is unpolarized .
 \label{fig3}}
\end{figure}

\begin{figure}
\includegraphics{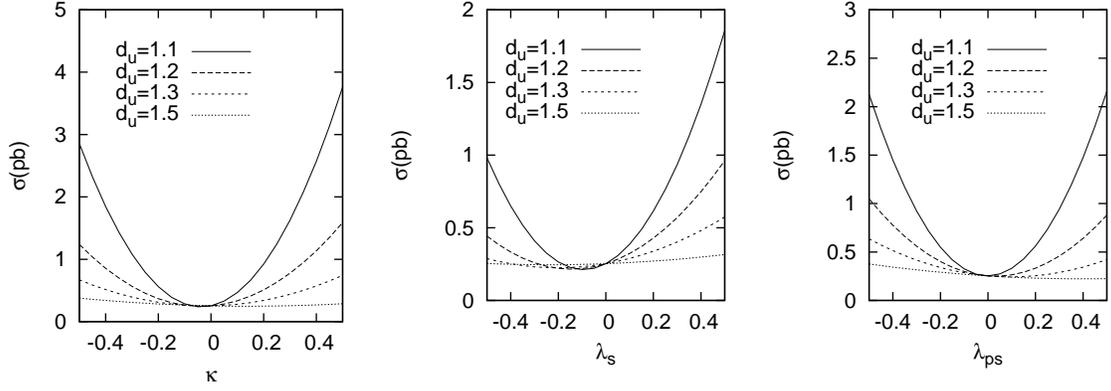}
\caption{ The same as Fig.3, but the top quark is in the right
helicity state. \label{fig4}}
\end{figure}

\begin{figure}
\includegraphics{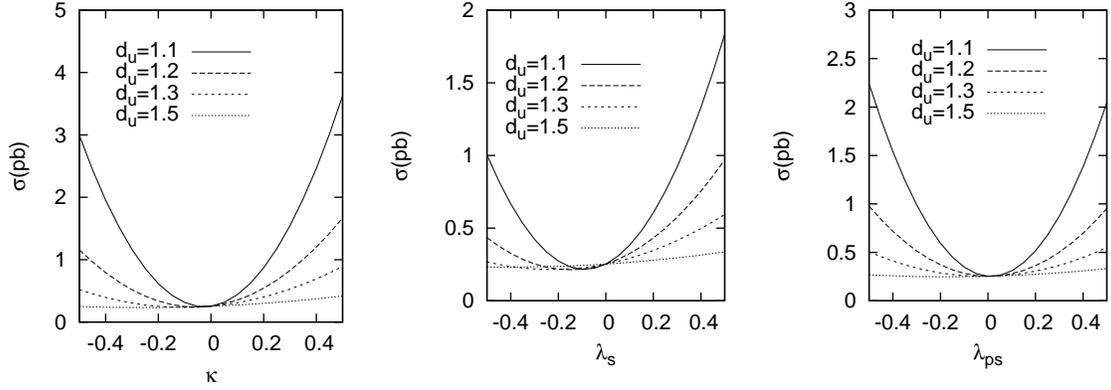}
\caption{ The same as Fig.4, but the top quark spin decomposition
axis is along one of the incoming photon beam direction. The spin
orientation of the top quark is spin up and the antitop quark is
unpolarized. \label{fig5}}
\end{figure}

\begin{figure}
\includegraphics{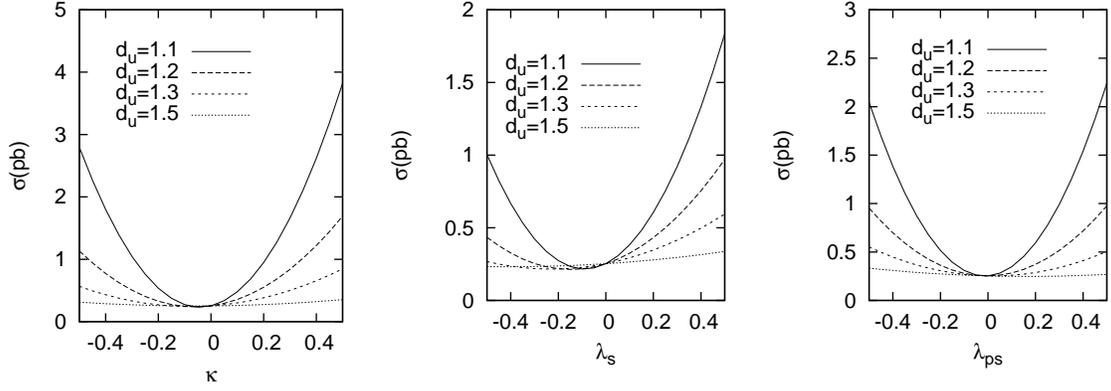}
\caption{ The same as Fig.5 but spin orientation of the top quark is
spin down.\label{fig6}}
\end{figure}

\begin{table} \caption{Sensitivity of $\gamma\gamma \to t\bar{t}$
to unparticle couplings at 95\% C.L. for various scale dimensions.
Top and antitop quarks are unpolarized. The center of mass energy of
the $e^+ e^-$ system is $\sqrt{s}=1$ TeV and $L_{int}=500 fb^{-1}$.
The unparticle energy scale is taken to be $\Lambda_{U}$=1
TeV.\label{tab1}}
\begin{ruledtabular}
\begin{tabular}{cccccc}
  &$d_U=1.1$&$d_U=1.3$&$d_U=1.5$&$d_U=1.7$&$d_U=1.9$ \\
\hline
 $\kappa$ &-0.049, 0.008 &-0.171, 0.015 &-0.458, 0.033 &-0.880, 0.067 &-0.794, 0.086 \\
 $\lambda_{S}$&-0.117, 0.009 &-0.413, 0.016 &-1.050, 0.034 &-2.062, 0.070 &-1.798, 0.092 \\
$\lambda_{PS}$&-0.025, 0.025 &-0.065, 0.065 &-0.158, 0.158 &-0.316, 0.316 &-0.341, 0.341\\
\end{tabular}
\end{ruledtabular}
\end{table}

\begin{table} \caption{ Sensitivity of $\gamma\gamma \to t\bar{t}$
to unparticle coupling $\kappa$ at 95\% C.L. for various spin
decomposition axes of the top and unpolarized antitop quark. The
center of mass energy of the $e^+ e^-$ system is $\sqrt{s}=1$ TeV
and $L_{int}=500 fb^{-1}$. The unparticle energy scale is taken to
be $\Lambda_{U}$=1 TeV.\label{tab2}}
\begin{ruledtabular}
\begin{tabular}{cccccc}
Spin Top &$\kappa$ $(d_U=1.1)$&$\kappa$ $(d_U=1.3)$&$\kappa$ $(d_U=1.5)$&$\kappa$ $(d_U=1.7)$&$\kappa$ $(d_U=1.9)$ \\
\hline \hline
$\gamma$-beam &    &    &  & &\\
 Up &-0.061, 0.009 &-0.222, 0.016&-0.566, 0.037 &-0.634, 0.132 &-0.282, 0.343\\
 Down &-0.090, 0.006 &-0.176, 0.021&-0.222, 0.095 &-0.795, 0.105 &-1.127, 0.086 \\
 \hline
Helicity &   &  &  & &\\
Right &-0.080, 0.007 &-0.084, 0.044 &-0.061, 0.344 &-0.069, 1.187 &-0.066, 1.453\\
Left &-0.076, 0.007 &-0.288, 0.013 &-0.702, 0.030 &-1.012, 0.083 &-0.483, 0.260\\
\end{tabular}
\end{ruledtabular}
\end{table}

\begin{table} \caption{ The same as Table \ref{tab2}, but for unparticle coupling  $\lambda_{PS}$.\label{tab3}}
\begin{ruledtabular}
\begin{tabular}{cccccc}
Spin Top &$\lambda_{PS}$ $(d_U=1.1)$&$\lambda_{PS}$ $(d_U=1.3)$&$\lambda_{PS}$ $(d_U=1.5)$&$\lambda_{PS}$ $(d_U=1.7)$&$\lambda_{PS}$ $(d_U=1.9)$ \\
\hline \hline
$\gamma$-beam &    &    &  & &\\
 Up &-0.019, 0.045 & -0.101, 0.059&-0.419, 0.084 &-1.132, 0.124 &-1.002, 0.165\\
 Down &-0.045, 0.019 &-0.059, 0.101 &-0.084, 0.419 &-0.124, 1.132 &-0.165, 1.002 \\
 \hline
Helicity &   &  &  & & \\
Right &-0.032, 0.028 &-0.028, 0.220 &-0.042, 0.839 &-0.078, 1.808 &-0.107, 1.533\\
Left &-0.028, 0.032 &-0.220, 0.028 &-0.839, 0.042 &-1.808, 0.078 &-1.533, 0.107 \\
\end{tabular}
\end{ruledtabular}
\end{table}

\begin{table}
\caption{Spin asymmetries for helicity basis and one of the incoming
photon beam direction. The unparticle couplings are taken to be
$\kappa$=1, $\lambda_S$=0 and $\lambda_{PS}$=1. The center of mass
energy of  the $e^+ e^-$ system is $\sqrt s$=1 TeV and
$\Lambda_{U}$=1 TeV. \label{tab4}}
\begin{ruledtabular}
\begin{tabular}{ccc}
Scale Dimension & $A_{\uparrow\,\downarrow}$\,\,(Helicity Basis) &$A_{\uparrow\,\downarrow}$\,\,($\gamma$ Beam Direction) \\
\hline \hline
$d_U$=1.1 &0.004 &-0.025 \\
$d_U$=1.3 &-0.157 &0.034 \\
$d_U$=1.5 &-0.342 &0.144 \\
$d_U$=1.7 &-0.274 &0.160 \\
$d_U$=1.9 &-0.197 &0.116 \\
\end{tabular}
\end{ruledtabular}
\end{table}

\end{document}